\documentclass[
 reprint,
 amsmath,amssymb,
 aps,
 prl, 
 floatfix,
 footinbib,
 sort&compress, numbers, merge
]{revtex4-2}

\usepackage{pdfpages}

\makeatletter
\AtBeginDocument{\let\LS@rot\@undefined}
\makeatother

\usepackage{color}
\usepackage[]{graphicx}
\usepackage{dcolumn}
\usepackage{bm}
\usepackage{float}
\usepackage{hyperref}
\usepackage{wrapfig}
\usepackage{comment}
\usepackage{ulem}

\renewcommand{\vec}[1]{\bm{\mathrm{#1}}}

\newcommand{\tripletcoord}{\vec{\xi}}

\graphicspath{{Figures/}}
\usepackage{tikz}
\usetikzlibrary{matrix}
\newcommand*\numcircled[1]{\tikz[baseline=(char.base)]{
            \node[shape=circle,draw,inner sep=1.2pt,scale=0.8,anchor=center] (char) {#1};}}

\newcommand{\fix}[1]{\noindent  {\colorbox{BrickRed}{\color{White} 
FIX:}  \color{BrickRed}#1\normalcolor}}

\begin{document}

\normalem

\title{Rapidly encoding generalizable dynamics in a Euclidean symmetric neural network: a Slinky case study}

\author{Qiaofeng Li$^{1,2}$}
\author{Tianyi Wang$^{2}$}
\author{Vwani Roychowdhury$^{2,*}$}
\author{M.K. Jawed$^{1,}$}
\email{V.R.: vwani@ee.ucla.edu, M.K.J.: khalidjm@seas.ucla.edu}
\affiliation{\footnotesize 
$^1$Dept.\ of Mechanical and Aerospace Engineering, University of California, Los Angeles, CA 90095, USA \\
$^2$Dept.\ of Electrical and Computer Engineering, University of California, Los Angeles, CA 90095, USA
}


\begin{abstract}
Slinky, a helical elastic rod, is a seemingly simple structure with unusual mechanical behavior; for example, it can walk down a flight of stairs under its own weight. Taking the Slinky as a test-case, we propose a physics-informed deep learning approach for building reduced-order models of physical systems. The approach introduces a Euclidean symmetric neural network (ESNN) architecture that is trained under the neural ordinary differential equation framework to learn the 2D latent dynamics from the motion trajectory of a reduced-order representation of the 3D Slinky. The ESNN implements a physics-guided architecture that simultaneously preserves energy invariance and force equivariance on Euclidean transformations of the input, including translation, rotation, and reflection. The embedded Euclidean symmetry provides physics-guided interpretability and generalizability, while preserving the full expressive power of the neural network. We demonstrate that the ESNN approach is able to accelerate simulation by one to two orders of magnitude compared to traditional numerical methods and achieve a superior generalization performance, i.e., the neural network, trained on a single demonstration case, predicts accurately on unseen cases with different Slinky configurations and boundary conditions.
%
\end{abstract}

\pacs{Valid PACS appear here}
\maketitle

\begin{figure*}[hbtp]
\includegraphics[width=0.95\textwidth]{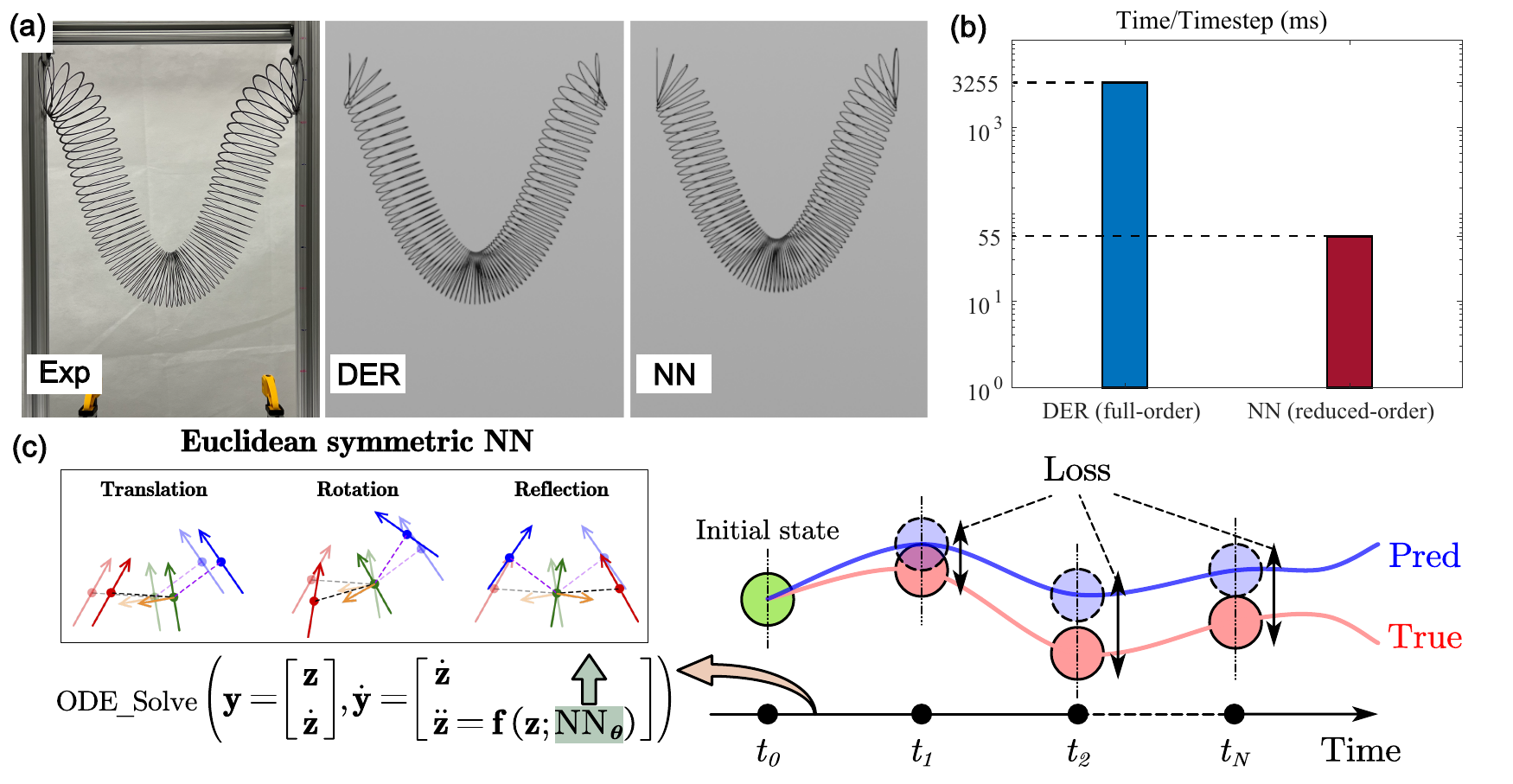}
\caption{(color online). (a) Static deformation comparison across a real-world Slinky experiment, the discrete elastic rod (DER) simulation, and the reduced-order model based on the Euclidean symmetric neural network. (b) Computation time comparison between DER simulation and the proposed NN method. (c) The core features of the proposed NN method: the NN embeds energy invariance and force equivariance on Euclidean transformations of the input Slinky configuration, including translation, rotation, and reflection. The NN is trained under the neural ordinary differential equation framework.}
\label{fig:Overall}  
\end{figure*}


\newcommand{\latentdynamics}{reduced-order dynamics}

Many dynamical systems are computationally expensive to simulate with classic numerical methods based on discretized differential equations, due to the fine spatiotemporal resolution required for the discretization to hold. More efficient yet effective dynamics predictions are widely demonstrated by humans. The high efficiency is largely due to their ability to make predictions at coarser resolution levels, sometimes for only perceivable states. %
Such simplification is similar to reduced-order modeling~\cite{bennerModelReductionApproximation2017,hartmanDeepLearningFramework2017,fultonLatentspaceDynamicsReduced2019,maulikTimeseriesLearningLatentspace2020} with a cognitive mapping from actual states to reduced (latent) states. The \latentdynamics{} describing the reduced degrees of freedom (DoFs) will consequently differ from first principles in physics, and need to be learned from observations with data-driven approaches.
An important tool that incorporates data-driven methods to augment physics models is deep learning \cite{Karniadakis2021Physics} based on deep neural networks (DNNs), universal approximators that show excellent data-fitting power.
On one hand, the data-driven models offer possibilities for superior computation efficiency compared to classic methods~\cite{hezavehFastAutomatedAnalysis2017,Fournier2020Artificial,heidenNeuralSimAugmentingDifferentiable2021}; e.g., as demonstrated in this Letter, the computation time can be cut down by one to two orders of magnitude.
On the other hand, it enables advances towards human-like, automated rule discovery and learning of the dynamics of a system from observations~\cite{Chua2019Reduced,rubanovaLatentODEsIrregularlySampled2019,championDatadrivenDiscoveryCoordinates2019,liFourierNeuralOperator2020,Iten2020Discovering,Liu2021Machine}.

Unconstrained neural networks (NNs), trained to directly fit time-series data generated by dynamical systems, however, often cannot replicate dynamics under unseen initial and boundary conditions because they fail to learn the underlying physics-constraints, such as Euclidean symmetry of space and conservation of energy. 
Indeed, recent works have shown that incorporating such constraints into deep learning is vital for generalization ability, learning efficiency, and interpretability~\cite{Zhang2018Machine,Atz2021Geometric}. For example, Hamiltonian Neural Network~\cite{Greydanus2019Hamiltonian} and Lagrangian Neural Network~\cite{Cranmer2020Lagrangian} introduced the physical prior of energy conservation to neural networks for learning dynamics.
Geometric deep learning \cite{Atz2021Geometric,bronsteinGeometricDeepLearning2021} and equivariant neural networks~\cite{weiler3DSteerableCNNs2018,thomasTensorFieldNetworks2018,finziPracticalMethodConstructing2021,satorrasEquivariantGraphNeural2021} leverage geometric symmetry properties of the network input to improve the quality of inherent knowledge learned by neural networks. Successful applications include drug discovery \cite{Gawehn2016Deep,Jimenez2021Artificial}, quantum chemistry \cite{Gilmer2017Neural}, and particle physics \cite{komiskeEnergyFlowNetworks2019,quParticleNetJetTagging2020}. 

In this Letter, we introduce a physics-informed deep learning approach to learning a 2D \latentdynamics{} of a Slinky. 
Instead of a long elastic helix in 3D space, a Slinky is described as a series of connected bars in a 2D plane \cite{holmesEquilibriaInstabilitiesSlinky2014}. 
The approach is guided by two principles: (i) The system trajectories along time are constructed by integrating the ordinary differential equation (ODE) formulated by Newton's second law of motion; and (ii) A neural network is trained to predict the 2D surrogate forces on the bars so that the observed trajectories can be generated following the ODEs. 
We endow the neural force predictor with Euclidean symmetry and energy conservation, which combines expressiveness with physics-guided generalizability.
In addition, the neural force predictor is faced with the challenge of compounding error, i.e., the error in force prediction at each time step can accumulate, leading to an erroneous simulated trajectory that seriously deviates from reality. We address this by going beyond learning separated state-force pairs, and use the Neural Ordinary Differential Equation (NODE) framework~\cite{Chen2018NODE} to match the entire trajectory. Under such framework, the consequence of a force prediction error in later time steps is considered in the loss, encouraging the neural network to be prescient and not to overfit proximate dynamics.

We show that this approach can accurately predict the motion of a real-world Slinky using 2D reduced-order representation of the Slinky. This approach is able to generalize to unseen Slinky configurations and boundary conditions by learning from a single demonstration case. The computational speed is increased by roughly 60 times compared to the traditional 3D modelling technique due to the significantly reduced DoFs, in spite of the relatively more complicated latent dynamics (Fig.~\ref{fig:Overall}). Moreover, the approach can generalize by exploiting ``physics" -- the neural force vector can be scaled with the material stiffness to predict the motion of a softer (previously unseen) Slinky. To the best of the authors' knowledge, we for the first time demonstrate, with experimental validation, that a deep learning approach is capable of extensively generalizing from one single learning case, and that a complex system in the real world could benefit from this reality-virtual-reality closed-loop pipeline.


\begin{figure}[htbp]
    \centering
    \includegraphics[width=0.48\textwidth]{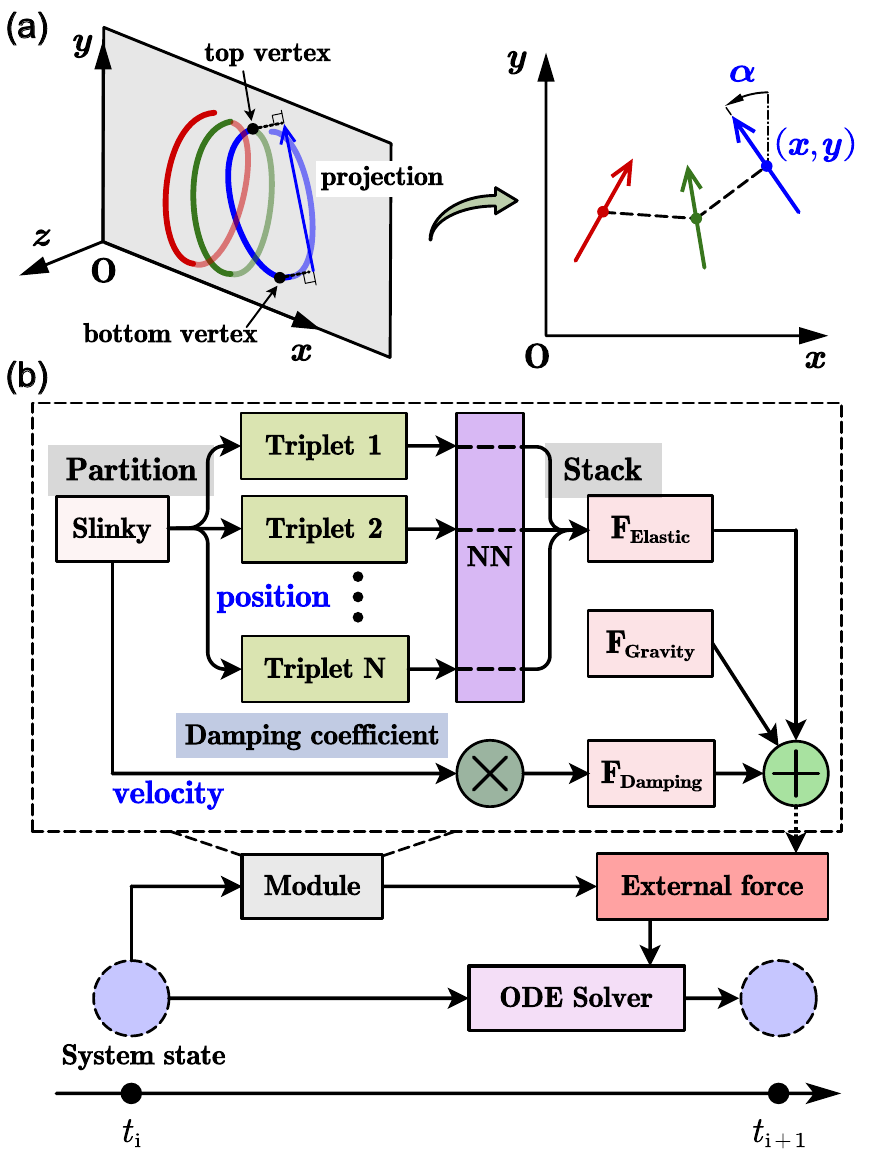}
    \caption{(color online). (a) Construction of the 2D bars by projecting 3D Slinky cycles into the XOY plane. Each 2D bar has 3 DoFs: $x$ and $y$ at its center, and the angle $\alpha$ with respect to the $y$-axis. (b). Time marching under the NODE framework. For the system at time $t_i$, the Slinky system is partitioned into triplets (3 adjacent bars) and passed through the same ESNN. The output vectors are stacked as the elastic force vector for the Slinky system. The total external force vector is the summation of elastic, gravity, and damping force vectors. The total external force and the system state at $t_i$ are fed into an ODE solver to update the system state at $t_{i+1}$.}
    \label{fig:NODE}
\end{figure}

\begin{figure*}[htbp]
\includegraphics[width=0.99\textwidth]{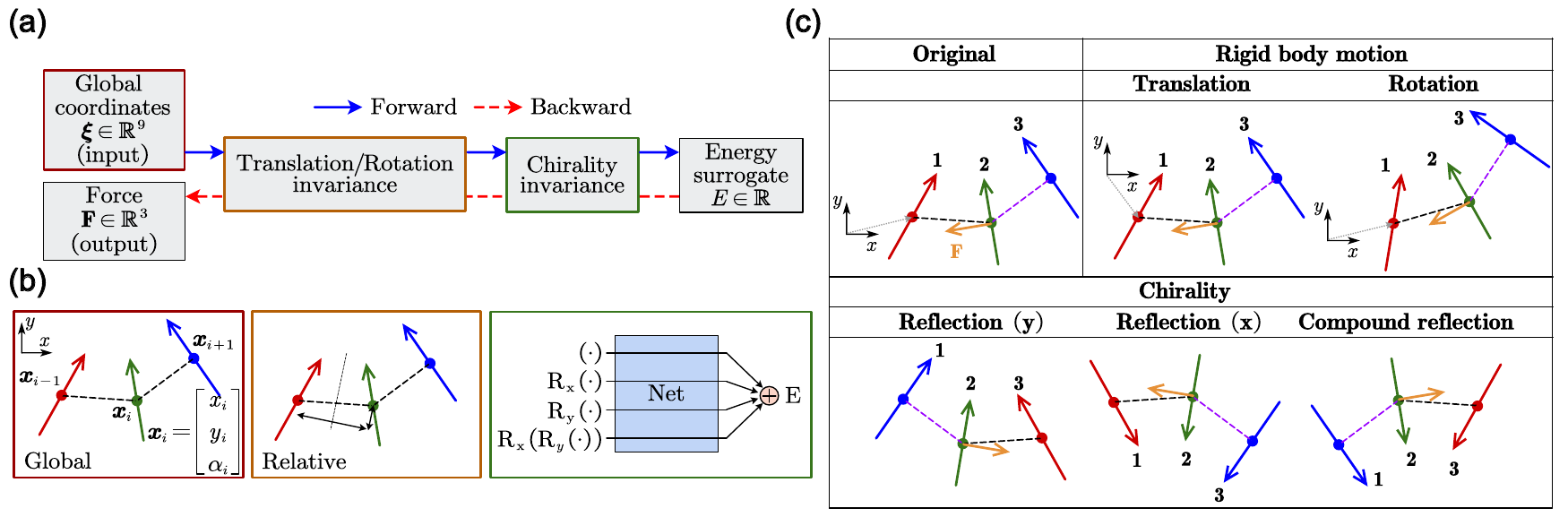}
\caption{(color online). (a) Workflow of the ESNN. The ESNN takes the global coordinates of a triplet as input, and generates the surrogate elastic force on the middle bar of the triplet as output. The ESNN first removes the rigid body DoFs from the input and then applies a chirality invariance module. An energy surrogate is calculated, which is invariant to rigid body and chiral transformations of the input. The output of the NN is computed by taking the derivative of the energy surrogate with respect to the coordinates of the middle bar in the input. The output is equivariant to rigid body and chiral transformations of the input. (b) Visualization of the input, rigid body motion removal, chiral transformation modules. The Net in the chiral transformation module is a DenseNet-like neural network of 5 hidden layers. For detailed neural network construction, refer to \cite{Supplement}. (c) Summary of the energy invariance and force equivariance properties of the ESNN. The original triplet configuration and its Euclidean transformed copies generate the same energy surrogate (invariance) and equivariant force vectors.}
\label{fig:GradientWorkflow}  
\end{figure*}

Given 3D Slinky motion data at a series of time steps, we first construct the 2D data as the ground truth for NN training. As shown in Fig.~\ref{fig:NODE}(a), the top and bottom vertices of each Slinky cycle are projected onto the XOY plane. The vector pointing from the projected bottom vertex to the projected top vertex of a cycle is the 2D representation of that cycle, referred to as a \emph{bar} in the following. The coordinates of the $i$th bar are $\mathbf{x}_{i}=[x_i,y_i,\alpha_i] \in \mathbb{R}^{3}$: $x_i$ and $y_i$ are the coordinates of the center point, and $\alpha_i$ is the angle with respect to the $y$ axis. Three adjacent bars group into a triplet with 9 coordinates. The surrogate elastic force on the middle bar of a triplet will be completely determined by these 9 coordinates. For a Slinky of $N_c$ cycles, its 2D representation has $N_c$ bars and $N_c$ triplets ($N_c$ instead of $N_c$-2 because of special boundary treatment), and the states of these $N_c$ bars at different time steps constitute the 2D system trajectory of the Slinky. Such a 2D reduced-order representation retains the essential geometry of the 3D structure of a Slinky, based on which a 3D helical shape is still reconstructable despite the substantial reduction in the number of DoFs. This property guarantees the reliable estimation of the elastic potential using 2D reduced-order states. See SM~\cite{Supplement} for boundary condition treatment and reconstruction of 3D Slinky shapes. After acquiring the 2D system trajectory, the reduced-order dynamics is learned by an ESNN under the NODE framework.

Fig.~\ref{fig:GradientWorkflow} shows the workflow of the ESNN. The input of the NN is the global coordinates of a triplet $\tripletcoord=[\mathbf{x}_{i-1}^{T}, \mathbf{x}_{i}^{T}, \mathbf{x}_{i+1}^{T}]^{T} \in \mathbb{R}^{9}$, where $\mathbf{x}_{i} \in \mathbb{R}^{3}$ is the coordinates of the $i$th bar and the superscript $T$ represents transpose operation. 
The first step in the ESNN is to make the representation invariant with respect to rigid body translation and rotation. This is accomplished by transforming the global coordinates $\tripletcoord$ into the relative coordinates  $\mathbf{z}\in\mathbb{R}^{6}$. Then $\mathbf{z}$ and its 3 reflected copies are passed through a DenseNet-like structure~\cite{Huang2017Densely} parameterized by $\boldsymbol{\theta}$, denoted as $f_{\boldsymbol{\theta}}(\cdot)$. The outputs are 4 scalars, and their summation is the energy surrogate $E$:
\begin{equation}
\begin{split}
    E(\mathbf{z}) = f_{\boldsymbol{\theta}}&(\mathbf{z}) + f_{\boldsymbol{\theta}}(R_{x}(\mathbf{z})) + \\ &f_{\boldsymbol{\theta}}(R_{y}(\mathbf{z})) + f_{\boldsymbol{\theta}}(R_{x}(R_{y}(\mathbf{z}))),
\end{split}
\end{equation}
where $R_x(\cdot)$ and $R_y(\cdot)$ stand for the reflection of the triplet coordinates about $x$ and $y$ axes. Note that the reflections are self-inverse and commutative, i.e.,
\begin{align*}
R_x(R_x(\cdot)) &= I(\cdot),\quad
R_y(R_y(\cdot)) = I(\cdot),~\textrm{and}\\
R_x(R_y(\cdot)) &= R_y(R_x(\cdot)),
\end{align*}
where $I(\cdot)$ is the identity transformation.
Therefore, the energy surrogate $E$ satisfies the following property:
\begin{equation}
    E(\mathbf{z}) = E(R_{x}(\mathbf{z})) = E(R_{y}(\mathbf{z})) = E(R_{x}(R_{y}(\mathbf{z}))).
\end{equation}
That is, $E$ is invariant to reflections on $\mathbf{z}$ and thus on $\tripletcoord$; $E$ is also invariant to rigid body transformation on $\tripletcoord$ due to such invariance of $\mathbf{z}$.
%
%
The output surrogate force vector $\mathbf{F}\in\mathbb{R}^{3}$ is generated by taking the derivative of $E$ with respect to $\mathbf{x}_i\in\mathbb{R}^{3}$ using the automatic differentiation mechanism in PyTorch~\cite{PyTorch2019Paszke}, i.e., $\mathbf{F}=\mathbf{F}(\tripletcoord)=\mathrm{\partial}E(\tripletcoord)/\mathrm{\partial}\mathbf{x}_i$.
It is straightforward to prove that $\mathbf{F}$ is equivariant to rigid body
and chiral transformation on $\tripletcoord$, i.e., denoting translation, rotation,
and reflection on $\tripletcoord$ as $T(\tripletcoord)$, $Ro(\tripletcoord)$, and
$Rf(\tripletcoord)$, we have $\mathbf{F}(T(\tripletcoord))=\mathbf{F}(\tripletcoord)$,
$\mathbf{F}(Ro(\tripletcoord))=Ro(\mathbf{F}(\tripletcoord))$, and
$\mathbf{F}(Rf(\tripletcoord))=Rf(\mathbf{F}(\tripletcoord))$.

The entire ESNN can be
considered as a general function
$\mathbf{F}=\mathrm{ESNN}_{\boldsymbol{\theta}}(\tripletcoord)$, while the aforementioned invariances / equivariances are
guaranteed regardless of the parameters $\boldsymbol{\theta}$. Conventionally, to obtain a good performance with an unconstrained neural network for all
orientation / chirality of the input, the approach is to augment
the training dataset with various rotation and reflection. In comparison, the
advantages of ESNN are trifold: (1) The Euclidean symmetry is strictly
guaranteed (instead of only approximated in the dataset-augmentation approach);
(2) Training cost is significantly lower due to the concise training dataset
without symmetry augmentation; (3) A compact NN architecture is possible since
weights are dedicated to learn the functional form of the surrogate elastic energy instead of the symmetries.
This bolsters the succeeding gain in computational speed. The entire workflow and the invariance / equivariance properties are illustrated in Fig.~\ref{fig:GradientWorkflow}.

Fig.~\ref{fig:NODE}(b) shows the schematic of the 2D reduced-order dynamics in our framework, specifically, how a 2D state is evolved from time $t_i$ to time $t_{i+1}$. At each time step, we first construct $N_c$ triplets and pass them through the ESNN. The output will be $N_c$ 3-dimensional vectors
representing the predicted surrogate elastic forces on the middle bars of the
triplets. These vectors are then concatenated to form the elastic force vector
for the entire Slinky system. The summation of this elastic force vector, the
gravity force vector, and the damping force vector is the total external force vector of the Slinky system. Then the system state can be updated by any ODE solver, 5th order Dormand–Prince method in this Letter. By repeating this update
procedure from a known initial state, a predicted trajectory for the
reduced-order Slinky system is calculated. The ESNN weights
$\boldsymbol{\theta}$ will be trained on the loss comparing the ground truth trajectory and the predicted one (Fig.~\ref{fig:Overall}(c)). This NODE training scheme enables accuracy and stability: the ESNN should not only predict surrogate forces accurately but also accurately in the way that after passing the predicted forces through an ODE solver, the errors in the calculated trajectories do not compound.

\begin{figure*}[htbp]
    \centering
    \includegraphics[width=0.95\textwidth]{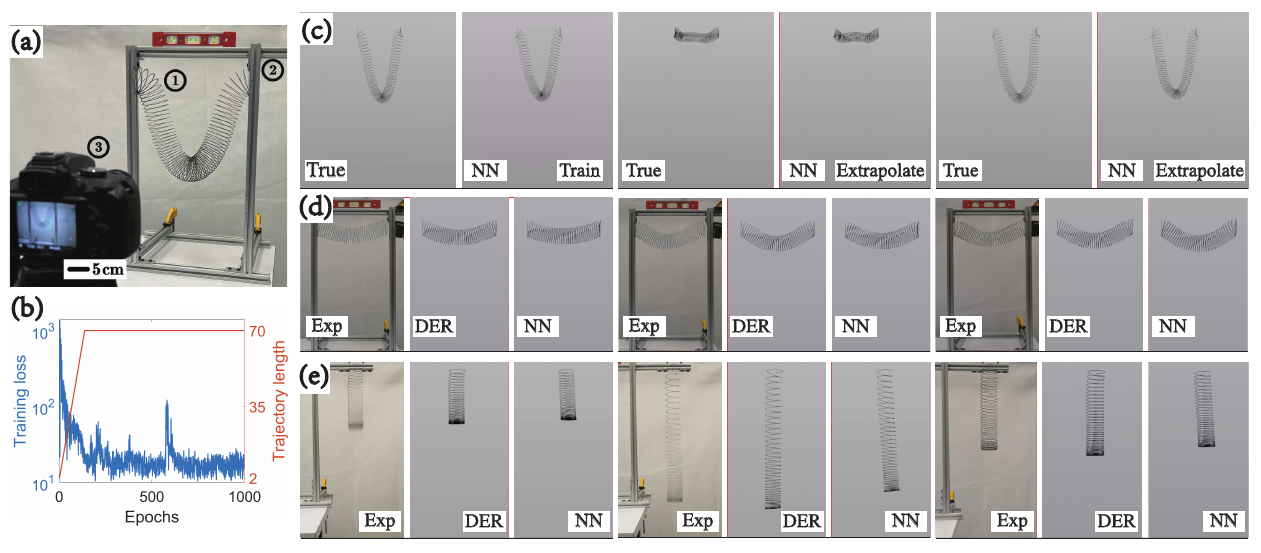}
    \caption{(color online). (a) Experimental apparatus: a 76-cycle Slinky \protect\numcircled{1} is supported by an experiment frame \protect\numcircled{2}. A camera \protect\numcircled{3} records the Slinky motion. (b) The loss history of the NODE training for 1000 epochs. The training trajectory length is gradually increased from 2 to 70. (c) Comparison between 3D DER ground truth and reconstruction from the ESNN reduced-order model in train phase and extrapolation phase. Time shots are at 0.45 s, 2.03 s, and 2.50 s (from left to right). (d) Comparison across real-world Slinky experiment, 3D DER simulation, and reconstruction from the ESNN reduced-order model. Testing is performed on a 40-cycle Slinky, unseen by the ESNN. Time shots are at 0.53 s, 1.43 s, and static (from left to right). (e) Comparison of testing on a 40-cycle Slinky with a different boundary condition and orientation from the training case. Time shots are at 0.2 s, 0.53 s, and static (from left to right).
    }
    \label{fig:Results}
\end{figure*}

We test the ESNN on a commercially-available 76-cycle Slinky (Poof-Slinky, Inc.). The experiment setup is shown in Fig.~\ref{fig:Results}(a). We record a single Slinky motion case using a camera, then calibrate a 3D discrete elastic rod (DER) model \cite{Bergou2008Discrete,Jawed2014Coiling,Jawed2015Propulsion,Huang2020Dynamic,Huang2020Shear} for comparison and generation of 2D training data, and then train the ESNN on the 2D data of this single case. After training, we freeze the NN model and test its generalization ability across several unseen cases.

First we measure the geometry and mass of the 76-cycle Slinky. Then a motion video is captured with its both ends clamped to the frame. The Slinky is initially held horizontal and dropped freely under gravity. A 3D DER model of the Slinky is built with the same geometry and mass. The Young's modulus and the damping parameter are calibrated to match the DER-generated Slinky motion with the video-recorded one. A 3D simulation is generated based on the calibrated model with the same initial and boundary conditions as the experiment and with the damping removed. The 2D system trajectory will be constructed based on projection of the generated 3D data and used to train an ESNN. The ESNN will be frozen after training. A damping and a contact model \cite{Li2020IPC} will be added in the NN deployment to match the real-world energy dissipation and non-penetration. The ESNN is trained on this one case, and tested for other cases with different boundary conditions, Slinky orientation, and number of cycles. For parameter calibration details, please refer to \cite{Supplement}.

When the ESNN is randomly initialized, a long trajectory prediction will consume too much ODE solution time. A better strategy is to make short trajectory predictions initially, update the ESNN weights, and then gradually increase the learning trajectory length. We start with trajectory length of 2 and increase the length by 1 every 2 epochs until it reaches 70. After that the trajectory length stays the same. The simulation time step is 0.01 s. So eventually the ESNN is trained on a ground truth trajectory of 0.7 s. In each epoch, the states of 20 randomly selected bars are used in the loss function:
\begin{equation}
    loss = \sum_{t_k} \sum_{i=1}^{20} \Vert \mathbf{W} (\mathbf{x}_{s_i}^{(k)} - \hat{\mathbf{x}}_{s_i}^{(k)}) \Vert_2^2 + \Vert \mathbf{W} (\mathbf{v}_{s_i}^{(k)} - \hat{\mathbf{v}}_{s_i}^{(k)}) \Vert_2^2
\end{equation}
where $\mathbf{W}\in \mathbb{R}^{3\times3}$ is a weight matrix, $s_i$ is the index of the $i$th randomly selected bar, $\mathbf{x}_{s_i}^{(k)}$ and $\mathbf{v}_{s_i}^{(k)}$ are the ground truth position and velocity of the $i$th randomly selected bar at time $t_k$, $\hat{\mathbf{x}}_{s_i}^{(k)}$ and $\hat{\mathbf{v}}_{s_i}^{(k)}$ are the corresponding predications. The training loss history and trajectory length are shown in Fig~\ref{fig:Results}(b).

After 1000 training epochs, a simulation is run with the trained ESNN for 2.5 s. The predicted trajectory, of which the first 0.7 s span is called train phase and the 0.7 - 2.5 s span extrapolation phase, is compared with 3D DER simulation in Fig.~\ref{fig:Results}(c). The ESNN results not only match the ground truth well within the train phase, but also perform well in the extrapolation phase up to 2.5 s, i.e., more than 2 times the training time span. The final static deformations for experiment, DER, and ESNN are in good agreement, as shown in Fig.~\ref{fig:Overall}(a). The computation times of one-second-long simulations for DER and ESNN are compared in Fig.~\ref{fig:Overall}(b). ESNN outperforms the traditional DER method by roughly 60 times due to the significant reduction in the number of DoFs. With the ESNN weights and all physical parameters for both DER and NN fixed, we test the generalization of the ESNN on previously unseen cases: on a Slinky (1) of a different number of cycles; (2) under a different boundary condition; (3) of a different density; (4) of a different Young's modulus (refer to \cite{Supplement} for (3) and (4)). In the first test case, the Slinky cycle number is changed to 40 with both ends clamped to the frame, as shown in Fig.~\ref{fig:Results}(d). The 40 cycle Slinky is initially held horizontal and then dropped freely under gravity. The dynamic motion at 0.53 s, 1.43 s, and the static deformation of the Slinky from experiment, DER simulation, and ESNN simulation are compared in the Fig.~\ref{fig:Results}(d) and showed an excellent agreement. In the second test case, the 40-cycle Slinky is held vertically, clamped at the upper end, and then dropped freely from its undeformed configuration. The dynamic motion at 0.2 s, 0.53 s, and the static deformation of the Slinky from experiment, DER simulation, and ESNN simulation are compared in the Fig.~\ref{fig:Results}(e) and again we observe a good agreement. In these two test cases, a satisfactory agreement is achieved without making any changes to the ESNN. The agreement originates from the embedded Euclidean symmetry of the NN and the fact that the NN is learning generalizable physics with a local construction. As a contrast, NN methods from \cite{liFourierNeuralOperator2020,Haghighat2021Physics} construct surrogate models for full-field solutions under a certain boundary condition. The benefit is that the NN prediction is extremely fast since only one forward-pass through the trained NN is required for each prediction. The price paid is in the generalization ability. For a different type of boundary condition, the NNs need to learn from scratch, and a new training dataset dedicated to the new boundary condition is required. However, in our ESNN approach, different boundary conditions can be readily incorporated after trained on a single case since the ESNN focuses on learning local physics which is boundary condition agnostic. In our second test case, the orientation of the Slinky is shifted by 90 degrees. The ESNN is still capable of generating the correct elastic forces and system trajectory prediction since it preserves rotation equivariance.

We have proposed an ESNN-based approach for building data-driven reduced-order models of physical systems under the NODE framework. We have validated its accuracy and extensive generalization ability, a critical differentiation from existing data-driven methods, on real-world Slinky experiments. A roughly 60 times computational acceleration is achieved compared with classic simulation methods. The generalization ability originates from the physics-guided design of the NN architecture, which embeds Euclidean symmetry, including translation, rotation, and chirality invariance for surrogate energy and equivariance for surrogate force. With these features, the ESNN is able to learn the reduced-order physics from a single demonstration case, and perform accurate and accelerated predictions on a variety of unseen cases. By incorporating only a widely applicable geometric symmetry instead of any domain specific knowledge or physical constraints, the ESNN possesses the promise to be universal to a wide variety of physical systems.

We thank Mingjian Lu for his assistance on simulation implementation, and Zhuonan Hao and Dezhong Tong for their assistance on experiments. We are grateful for financial support from the National Science Foundation (NSF) under award number CMMI-2053971. Q.L. and M.K.J are grateful for support from NSF (IIS-1925360). M.K.J is grateful for support from NSF (CAREER-2047663, CMMI-2101751).

\hypersetup{hidelinks}

\providecommand{\noopsort}[1]{}\providecommand{\singleletter}[1]{#1}%

\clearpage

\onecolumngrid 


\section*{Supplementary material}

# Rapidly encoding generalizable dynamics in a Euclidean symmetric neural network: a Slinky case study
## – Supplemental Information –

Qiaofeng Li, Tianyi Wang, Vwani Roychowdhury, and M.K. Jawed

## S1 Discrete Elastic Rod simulation

The dynamics of thin rods is simulated with the Discrete Elastic Rod (DER) method. A rod is discretized into $M$ vertices along its centerline in the 3D space. The vertices are denoted as $\mathbf{x}_0, \ldots, \mathbf{x}_{M-1}$, where $\mathbf{x}_i \in \mathbb{R}^3$. $M-1$ edge vectors, $\mathbf{e}^0, \ldots, \mathbf{e}^{M-2}$, are formed by $\mathbf{e}^i = \mathbf{x}_{i+1} - \mathbf{x}_i$. Here we use subscripts to denote quantities associated with vertices and superscripts when associated with edges. Each edge, $\mathbf{e}^i$, has two frames: an orthonormal reference frame denoted by $\{\mathbf{d}_1^i, \mathbf{d}_2^i, \mathbf{t}^i\}$ and a material frame denoted by $\{\mathbf{m}_1^i, \mathbf{m}_2^i, \mathbf{t}^i\}$. $\mathbf{t}^i = \mathbf{e}^i/|\mathbf{e}^i|$ is the unit tangent vector for the $i$th edge. The angle $\theta^i$ that rotates the reference frame to the material frame is the twisting angle of the $i$th edge. The vertex positions and edge twisting angles constitute the $4M-1$ degrees of freedom (DoFs) of the entire rod system, $\mathbf{q} = [\mathbf{x}_0^T, \theta^0, \mathbf{x}_1^T, \ldots, \mathbf{x}_{M-2}^T, \theta^{M-2}, \mathbf{x}_{M-1}^T]^T$, where the superscript $T$ stands for transposition.

The elastic energy of a rod is composed of three parts: stretching, bending, and twisting, based on the Kirchhoff's rod theory. For a rod with Young's modulus $E$, shear modulus $G$, a rectangular cross section $A$, bending moment of inertia $I_1$ and $I_2$, and twisting moment of inertia $J$, the stretching, bending, and twisting energies are given by [34] as

$$E_s = \frac{1}{2} \sum_{i=0}^{M-2} EA(\epsilon^i)^2 |\bar{\mathbf{e}}^i| \tag{1a}$$

$$E_b = \frac{1}{2} \sum_{i=0}^{M-1} \frac{E}{\Delta l_i} \left[ I_1 \left( \kappa_i^{(1)} - \bar{\kappa}_i^{(1)} \right)^2 + I_2 \left( \kappa_i^{(2)} - \bar{\kappa}_i^{(2)} \right)^2 \right] \tag{1b}$$

$$E_t = \frac{1}{2} \sum_{i=0}^{M-1} \frac{GJ}{\Delta l_i} (\tau_i)^2 \tag{1c}$$

where $\Delta l_i = \left(|\bar{\mathbf{e}}^i| + |\bar{\mathbf{e}}^{i+1}|\right)/2$ is the Voronoi length of the $i$th vertex. $|\bar{\mathbf{e}}^i|$ and $\epsilon^i$ are the undeformed length and stretching strain of the $i$th edge.

$$\epsilon^i = \frac{|\mathbf{e}^i|}{|\bar{\mathbf{e}}^i|} - 1 \tag{1}$$

$\kappa_i^{(1)}$ and $\kappa_i^{(2)}$ are the material curvatures at the $i$th vertex,

$$\begin{aligned} \kappa_i^{(1)} &= \frac{1}{2} \left( \mathbf{m}_2^{i-1} + \mathbf{m}_2^i \right) \cdot (\kappa \mathbf{b})_i, \\ \kappa_i^{(2)} &= -\frac{1}{2} \left( \mathbf{m}_1^{i-1} + \mathbf{m}_1^i \right) \cdot (\kappa \mathbf{b})_i. \end{aligned} \tag{2}$$

$(\kappa \mathbf{b})_i$ is the bending strain at the $i$th vertex.

$$(\kappa \mathbf{b})_i = \frac{2 \mathbf{e}^{i-1} \times \mathbf{e}^i}{|\mathbf{e}^{i-1}||\mathbf{e}^i| + \mathbf{e}^{i-1} \cdot \mathbf{e}^i} \tag{3}$$

and its norm is $|(\kappa \mathbf{b})_i| = 2\tan(\phi_i/2)$. $\tau_i$ is the discrete twisting strain at the $i$th vertex,

$$\tau_i = \theta^i - \theta^{i-1} + m_i^{ref} \tag{4}$$

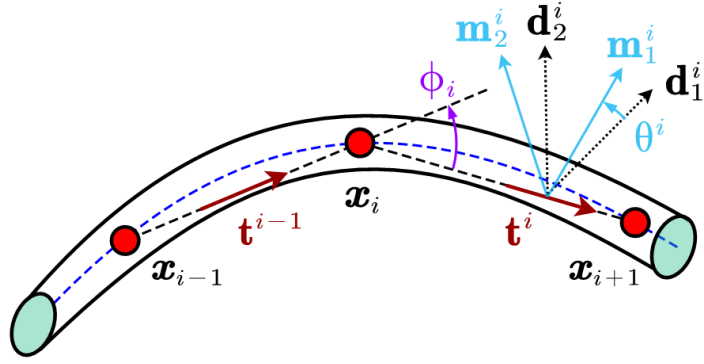

Figure S1: (color online). The schematic of a discrete elastic rod. For three adjacent vertices $\mathbf{x}_{i-1}$, $\mathbf{x}_i$, and $\mathbf{x}_{i+1}$, $\phi_i$ denotes the bending angle at $i$the vertex. $\mathbf{d}_1^i$ and $\mathbf{d}_2^i$ are the reference frame directors of the $i$th edge. $\mathbf{m}_1^i$ and $\mathbf{m}_2^i$ are the material directors. $\theta^i$ is the twisting angle for the $i$th edge.

where $m_i^{ref}$ is the reference twist associated with the reference frame.

The schematic is shown in Fig. S1. The generalized elastic force on the system is calculated as

$$\mathbf{F}^e = -\frac{\partial(E_s + E_b + E_t)}{\partial \mathbf{q}} \quad (5)$$

By treating the rod system as an equivalent mass-spring system, we can update the system position $\mathbf{q}$ by solving the following governing equation

$$\mathbf{M}\ddot{\mathbf{q}} - \mathbf{F}^e = \mathbf{0} \quad (6)$$

where $\mathbf{M}$ is the mass matrix of the system. In this Letter, the above governing equation is solved by the implicit Euler method.

## S2 Incremental Potential Contact model

In order to enforce the non-penetration condition between Slinky cycles, we adopt the Incremental Potential Contact model [39] to simulate the contact forces.

The total contact energy of the system is

$$E_c = \lambda \sum_{k \in \mathbb{C}} b(d_k) \quad (7)$$

where $\lambda$ is a Lagrangian multiplier. $\mathbb{C}$ is the set of all possible pairs of edges in the DER-discretized Slinky system. $d_k$ is the distance between the edges in the $k$th edge pair. $b(\cdot)$ is the barrier function

$$b(d, \hat{d}) = \begin{cases} -(d - w - \hat{d})^2 \ln\left(\frac{d-w}{\hat{d}}\right), & 0 < d - w < \hat{d} \\ 0 & d - w \geq \hat{d} \end{cases} \quad (8)$$

where $\hat{d}$ is the control parameter for the barrier function. $w$ is the width of the Slinky cross section. In this Letter, $\lambda = 10^4$, $\hat{d} = 0.1$ mm, and $w = 0.339$ mm. The advantages of this barrier function are: (1) The barrier is 2nd-order continuous with respect to $d$, which improves simulation convergence; (2) The barrier function tends to infinity as $d - w \to 0$ regardless of the Lagrangian multiplier value $\lambda$. This means that theoretically no penetration is allowed no matter how large the relative velocity is between two edges. This marks a significant difference from traditional contact models where the contact energy barrier is capped at a certain value so that penetration is inevitable when the relative velocity is too large between two edges.

$d_k$ is a function of the system DoFs $\mathbf{q}$. So the contact force vector is calculated as

$$\mathbf{F}^c = -\frac{\partial E_c}{\partial \mathbf{q}} \quad (9)$$

Taking damping (modelled as a viscous force proportional to velocity) and contact forces into consideration, the system governing equation eventually becomes

$$\mathbf{M}\ddot{\mathbf{q}} - \mathbf{F}^d - \mathbf{F}^e - \mathbf{F}^c = \mathbf{0} \tag{10}$$

## S3 The DenseNet-like NN structure

The "Net" part in the Euclidean symmetric neural network (ESNN) is a DenseNet-like structure [32], where shortcut pathways are created for a layer from all its previous layers. The Net receives a $\mathbb{R}^6$ vector as input. It first increases the feature dimension to 32 by a fully-connected (FC) layer. Then the feature is passed through FC layers with Softplus activation. A new feature with an increased dimension is formed by concatenating the previous feature with the FC layer output, i.e.,

$$\mathbf{x}_i = \begin{bmatrix} \mathbf{y} \\ \mathbf{x}_{i-1} \end{bmatrix} = \begin{bmatrix} \mathrm{FC}_{\mathrm{Softplus}}(\mathbf{x}_{i-1}) \\ \mathbf{x}_{i-1} \end{bmatrix} \tag{11}$$

After 5 layers, the feature dimension is increased to 192. This feature is then squared and passed through a FC layer with no activation to produce the final scalar output.

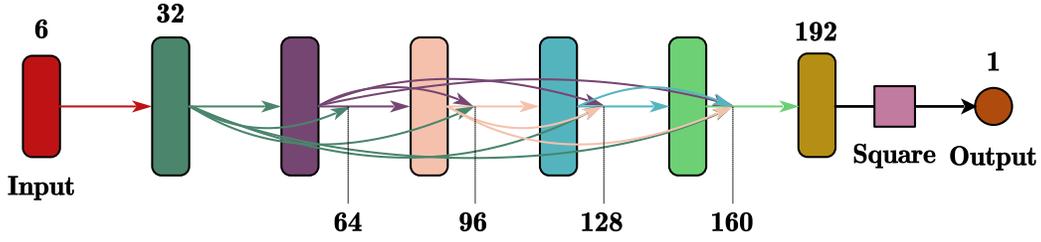

Figure S2: (color online). The DenseNet-like NN first increases the feature dimension from 6 to 32 by a FC layer without activation, and then in the subsequent 5 layers, concatenating the feature in a layer with the output of a FC layer to produce the feature for the next layer. This is equivalent to creating shortcuts for a layer from all of its previous layers. The feature dimension will increase by 32 after passing through each layer and eventually reach 192. Then the feature is squared and passed through a FC layer without activation to generate a scalar output.

## S4 Boundary condition treatment of 2D triplets

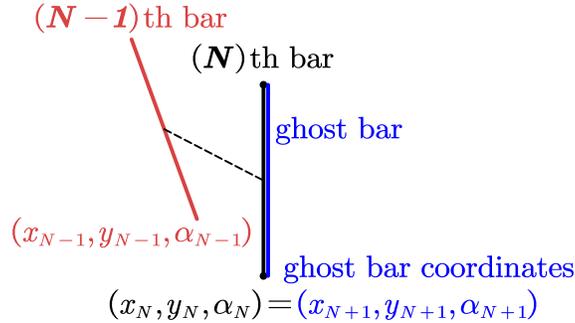

Figure S3: (color online). The boundary treatment for free end boundary conditions. The ghost bar (the $N+1$th bar) has the same coordinates as the $N$th bar.

For clamped boundary conditions, no special treatment is needed since it is unnecessary to calculate the elastic force on the boundary bar. For free boundary conditions, a ghost bar is constructed, as

3

shown in Fig. S3. Assume that the $N$th bar of a $N$-cycle Slinky is free. Then the coordinates of the ghost bar (the $N+1$th bar) are the same to those of the $N$th bar, i.e., $(x_{N+1}, y_{N+1}, \alpha_{N+1}) = (x_N, y_N, \alpha_N)$. The elastic force on the $N$th boundary bar is calculated with

$$(x_{N-1}, y_{N-1}, \alpha_{N-1}, x_N, y_N, \alpha_N, x_{N+1}, y_{N+1}, \alpha_{N+1})$$

as the ESNN input.

## S5  3D Slinky reconstruction

The reconstruction from 2D bar to 3D helical shape of the Slinky is shown in Fig. S4. Given the positions of a bar, a half-circle of radius $R$ is drawn perpendicular to the XOY plane, from the top vertex to the bottom vertex of the bar. $R(=33$ mm in our experiment) is the Slinky helical radius at its initial undeformed shape. Another half-circle is drawn from the bottom vertex of the bar to the top vertex of the next bar. These two half-circles form one Slinky cycle. Repeating this process for all bars, a helical Slinky shape is drawn.

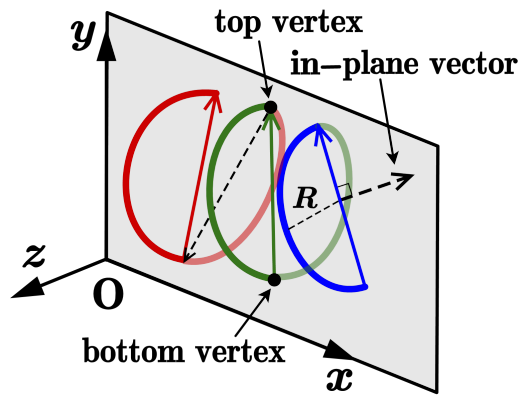

Figure S4: (color online). Reconstruction from 2D reduced-order model to 3D helical shape. For each bar, two half circles are drawn to form a Slinky cycle: one from the top vertex to the bottom vertex of the bar, the other from the bottom vertex of the bar to the top vertex of the next bar. The half circles are perpendicular to the XOY plane.

## S6  The experiment calibration procedure

The dimensions of the 76-cycle Slinky are shown in Fig. S5. The mass of the Slinky is 192.8 g. The volume is $2.1357 \times 10^{-5}$ m$^3$. The density is $9.0 \times 10^3$ kg/m$^3$. In order to calibrate the Young's modulus, we perform static deformation tests on the Slinky and match the 3D DER simulation results to the experiment results under the same geometric parameters and boundary conditions. The final calibrated Young's modulus is 69 GPa. The experimental static deformation of the Slinky and the DER simulation result under the vertical hanging condition are compared in Fig. S5(d) and a good matching is observed.

## S7  Generalization to different density and elasticity

In the first generalization test case, the density of the Slinky is halved to $4.5 \times 10^3$ kg/m$^3$ in both DER and NN simulation. The Young's modulus is 69 GPa. The rest of the parameters remain the same. The Slinky is clamped at both ends and dropped freely under gravity. The simulation results are shown in Fig. S6. In the second generalization test case, the Young's modulus of the Slinky is halved to 34.5 GPa in both DER and NN simulation. The density is $9.0 \times 10^3$ kg/m$^3$. The rest of the parameters and the boundary conditions remain the same. The results are shown in Fig. S7.

4

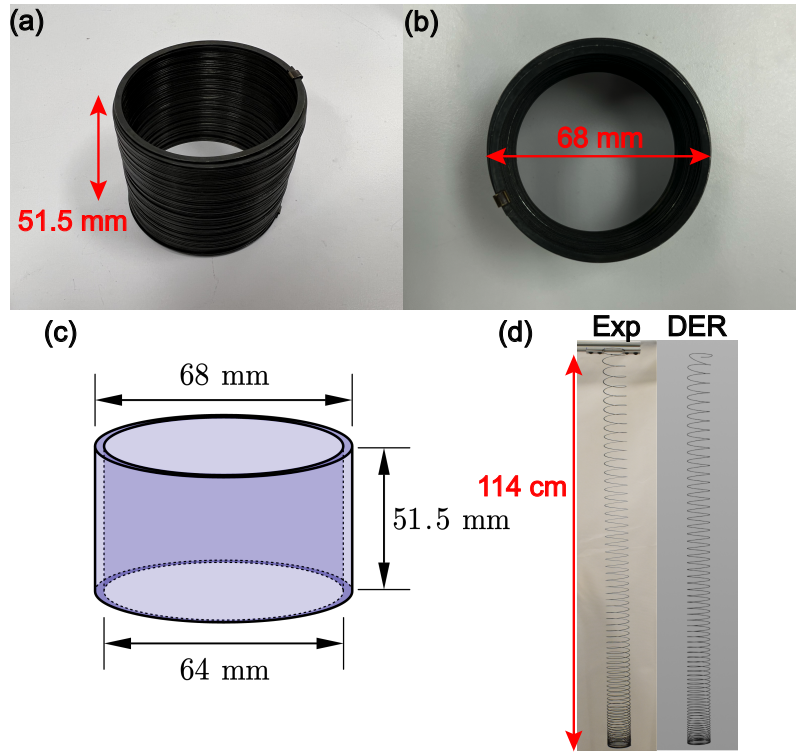

Figure S5: (color online). The side (a) and top (b) view of the Slinky at its undeformed condition. (c) When calculating the volume of the Slinky, it is treated as a cylinder of outer diameter 68 mm, inner diameter 64 mm, and height 51.5 mm. (d) The comparison of static deformation of the Slinky between experiment and DER under the vertical hanging condition.

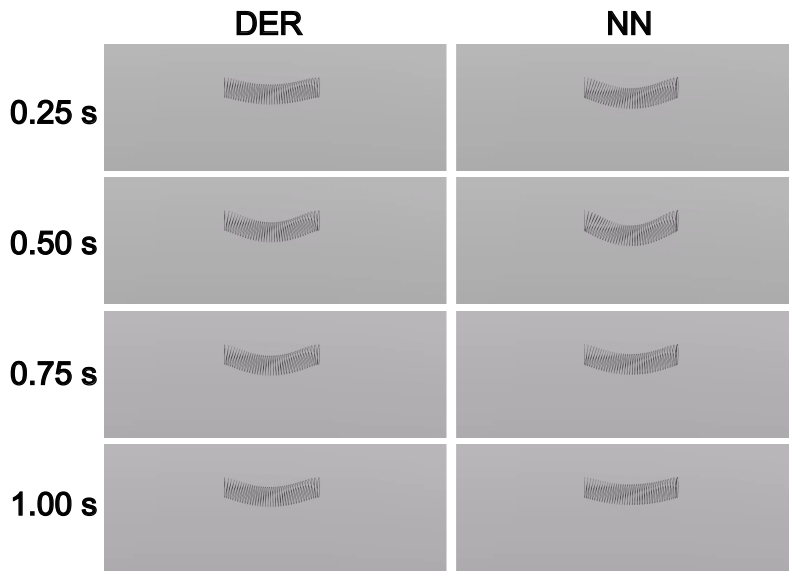

Figure S6: (color online). The dynamic motion of the Slinky with a halved density $4.5 \times 10^3$ kg/m$^3$ and a Young's modulus 69 GPa at different time steps from DER and NN simulation.

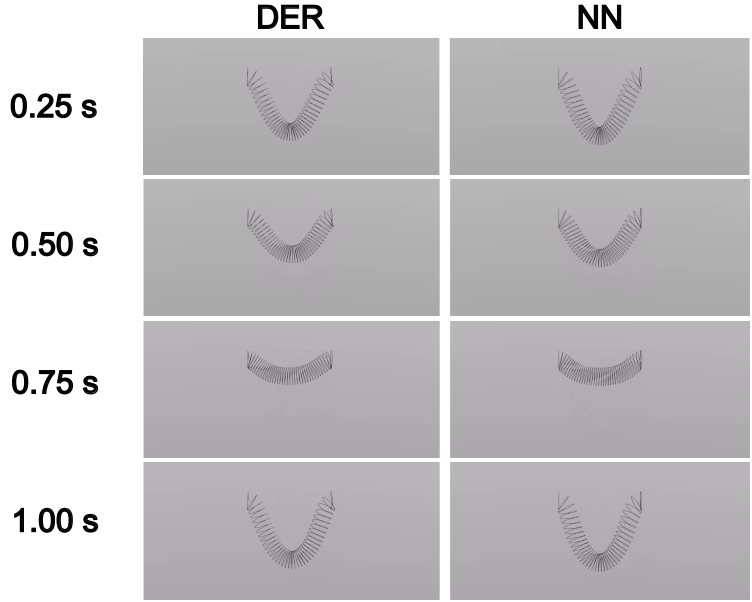

Figure S7: (color online). The dynamic motion of the Slinky with a density $9.0 \times 10^3$ kg/m$^3$ and a halved Young's modulus 34.5 GPa at different time steps from DER and NN simulation.

## S8 Simulation environment

3D DER simulation is performed in C++ with an in-house DER code base. The NN training is performed in the PyTorch framework [33]. After training, the NN is saved, imported into the in-house C++ code base, and executed using the LibTorch library. The 2D reduced-order simulation only replaces the elastic force calculation in 3D DER with the NN calculation. The damping model, contact model, and the rest of modules are the same for 3D and 2D simulation. The performance is tested on a personal computer with a 4-core CPU and 8 GB RAM.

## S9 Supplemental movie

The supplemental movie (https://youtu.be/mHCFa8U9Xpw) corresponds to one training case and four testing cases. They are described below.

1. Training case (first 0.7 s training phase, later than 0.7 s extrapolation phase).

2. Keeping all the other parameters fixed, the number of Slinky cycles is reduced to 40.

3. The orientation of the Slinky is changed from horizontal to vertical. The bottom end boundary condition is changed from clamped to free end.

4. Keeping all the other parameters fixed, we changed the density from $9 \times 10^3$ kg/m$^3$ to $4.5 \times 10^3$ kg/m$^3$.

5. Keeping all the other parameters fixed, we changed the Young's modulus from 69 GPa to 34.5 GPa.

6



\end{document}